\documentclass[aps,superscriptaddress]{revtex4}
\usepackage{graphicx,psfrag,amsmath,times,hhline}
\usepackage{epsf}
\usepackage{epsfig}
\newcommand{\beeq}{\begin{equation}}
\newcommand{\eneq}{\end{equation}}
\newcommand{\beeqar}{\begin{eqnarray}}
\newcommand{\eneqar}{\end{eqnarray}}
\newcommand{\be}{\begin{equation}}
\newcommand{\ee}{\end{equation}}
\newcommand{\br}{\begin{eqnarray}}
\newcommand{\er}{\end{eqnarray}}

\def\keywords{\vspace{.5em}
{\textit{Keywords}:\,\relax%
}}
\def\endkeywords{\par}
\begin{document}

\title{Ontology and Quantum Mechanics}
\author{N.D.~Hari Dass}
\email{dass@cts.iisc.ernet.in}
\affiliation{Tata Inst. for Fundamental Research,TCIS, Hyderabad}
\affiliation{Chennai Mathematical Institute,Chennai, India
.}
\affiliation{CQIQC, IISc, Bangalore, India
.}
\begin{abstract}
The issue of ontology in quantum mechanics, or equivalently the issue of the reality of the wave function is
critically examined within standard quantum theory. It is argued that though no strict ontology is possible within quantum theory,
ingenious measurement schemes may still make the notion of a \emph{FAPP Ontology} i.e ontology for all practical purposes (a phrase
coined by John Bell), meaningful
and useful. 
\end{abstract}
\vspace{.5em}
\maketitle


\endkeywords
\section{Introduction}
A cursory check as to the meaning of the word \emph{Ontology} will turn up a bewildering response, with a wide spectrum of interpretations.
So is also the case for its close relative \emph{Epistemology}. It is not the purpose of this article to get into a general discourse on
this concept. Instead, it will focus on its meaning as widely understood by physicists, more particularly the quantum physicists. 
Though notions of existence and of reality come freqently associated with ontology, we shall focus more on aspects of reality. In the
specfic context of quantum theory, this more or less concerns the so called reality of the wavefunction. Reality is in itself a heavily loaded
concept were one to turn into it from general philosophical considerations. We shall therefore restrict attention to \emph{Physicist's
notion of reality}, however unsophisticated it may appear to philosophers at large!

It is fair to say that the notion of reality to most physicists is conditioned by their experience from classical physics. Many so called
paradoxes in quantum theory have in fact arisen because of this. Nevertheless, a careful examination of the concept of reality in classical
physics is essential as a guide to examining its counterpart in quantum theory. It is clear that even in classical physics, notions of
reality are intimately tied up with aspects of observation, or of measurements. Therefore, the plan of this article is to first examine
ontology in classical physics, and to identify those aspects of classical measurements and dynamics that make the notion of reality reliable
and useful. We then examine the issue of ontology in quantum mechanics against the backdrop of a variety of quantum measurements all the
way from the Dirac-von Neumann description to the current day explosions.
\section{Ontology in Classical Physics}
Reality in classical physics may be characterised by certain \emph{robust} associations of \emph{attributes} and \emph{objects}. For example,
when one says that a particular Rose is Red, this represents an element of reality with many important aspects, many of which appear trivial
and straightforward unless carefully contemplated upon. In this case the attribute is Redness and the object is the Rose in question. What
are the mechanisms in classical physics that bring about this association, and in what sense this association is robust are questions whose
answers hold the key to a finer understanding of reality in classical physics.

Before attempting to answer them, let us expand the list of attributes in this case to include, let us say, \emph{Smell}. Classical reality
says that these two attributes can peacefully coexist and that the reality of one need not interfere with the reality of the other. Now what
gives an element of reality to, say, the redness, is that no matter how many times we observe the colour, no matter how we observe the
color, or no matter how often we interject color observations with other observations, say in this case smell, we come up with the
same measure of redness for the flower. It is obvious that this is possible only if observing the color of the rose does not itself alter
its color. 

We can sharpen this by introducing the notion of a \emph{state} of the object; let us stipulate that the state of any object is specified
by the \emph{values} of its attributes. 
In this example, red is the 'value' of the attribute of color for this particular state of the rose
which may be called a 'red rose'. One could have yellow roses, purple roses etc and they would all refer to different states  the object; let us stipulate that the state of any object is specified
by the \emph{values} of its attributes. In this example, red is the 'value' of the attribute of color for this particular state of the rose
which may be called a 'red rose'. One could have yellow roses, purple roses etc and they would all refer to different states. 
It is worth recalling a characterization of a state by Dirac \cite{diracbook1,diracbook2}; though it was given in the context
of quantum theory, it is pertinent to any theory, and certainly to classical physics also. 
According to him, a state is an embodiment of all possible measurement outcomes.

At this point
there is an important subtlety that needs to be taken care of. It appears and in reality it is indeed so, that a red rose is a different
object from a yellow rose and we are not really talking of different states of an object but of different objects. In fact that 
may make the distinction between the object and its state artificial and unwarranted. To overcome that, we shall allow for the possibility
(not altogether unrealistic) of processes that could change the color of a given rose. Then the rose, the given object, can indeed be in different states of color. If, as mentioned above, we are also considering additional attributes like smell, a characterization of the state of a
rose would, in this classical context, require specifying the values of both smell and color. These values can be thought of as the outcomes
of color and smell measurements thus tying up with the characterization of a state according to Dirac.

In fact, we can go to the more prosaic world of classical mechanics and consider a particle as the object, its position, velocity etc.
as its attributes. The state of the particle is then specified by the values of these attributes. There is an obvious redundancy with this
description. For example, one could have also considered, for example, the square of the position as an attribute, but then that would not
carry any additional information from that already carried by the position on its own. So a distinction should be made between what one
may call \emph{primary} and \emph{derived} attributes. The upshot is that it is enough to consider an optimal set of independent attributes
for state description.

Let us return the issue of reality and its robustness. The association of, say, position with the particle can be taken as an element of
reality which is robust because measurement of position of the particle returns the same values within some range of errors(more on this later)
no matter how often this measurement is done, how this measureent is done(as there are many means of position measurements), and in what order
these measurements are done in the sense that position measurements may be interspersed with measurements of other attributes. In fact, in
the classical world it would then be possible to say that this element of reality exists even if no one is actually observing the particle!

It is obvious that this is possible if and only if the measurements have no effect on the state. Such measurements can be called 
\emph{non-invasive}. But it is obvious that not every measurement be necessarily non-invasive even in classical physics. One could in 
principle adopt a measurement scheme that is deliberately invasive. For example, a position measurement of a tiny particle could be done 
by hitting it with a big stone. So a choice of non-invasive measurement is essential in the scheme above. In the classical world, by and large
measurements are non-invasive unless by deliberate design.

It is important to emphasize that non-invasiveness by itself is enough to guarantee a robust element of reality. Now the other crucial aspect
of the classical world enters the picture and this is \emph{determinism}. To appreciate this consider the possibility that before a measurement
the particle is in a definite state i.e with definite values of all its attributes. A non-invasive measurement may leave the particle in
the same state, but may not necessarily yield definitive values for these attributes. This could happen when the physical processes making up
the measurement are not deterministic. It could well be that a definite measured value emerges upon averaging a large number of outcomes. 
Such a world would exhibit both ontic and epistemic features.

But the world of classical physics is deterministic. On top of that, no separate laws have to be stipulated for measurements. Therefore
in principle every classical measurement should yield definite outcomes with no errors at all. But errors do occur in classical measurements.
This is for the obvious reason that even in the deterministic world of classical physics, not every source of influence in an experiment can
be identified and accurately accounted for. A pragmatic approach would treat the unknowns \emph{probabilistically} thereby introducing
randomness in a pefectly deterministic world! Therefore the outcomes will have variances and actual errors can be statistically reduced through
repeated measurements.

Nevertheless, even this randomness introduced purely for practical purposes, governs errors that can be controlled by better experimental
designs. Then, one can adopt the reasonable stand that the outcomes within such narrow and controllable errors are, for all practical
purposes, making the strictly non-invasive measurements into practically non-invasive measurements.

But it is worth appreciating that any randomness, however small, does not allow for ontic descriptions, in principle. But in practice 
this does not pose a problem. In that sense, even the 'real' world of classical physics has a blurry edge, which we ignore all the time!

All these considerations have one profound consequence. Measurements on a single object are meaningful, and statistical errors can be
meaningfully reduced arbitrarily by making suffieciently large number of repeated measurements on the same object. It should be stressed
that this arose both due to the near non-invasive measurements as well as due to each measurement practically yielding full information.

Even with regard to deliberately invasive measurements, the determinism of classical physics can in principle, though tedious and
heavy on resources in practice, provide a means of compensating for the invasive effects. In the example of throwing a rock to measure
the position of a small particle, though the rock strongly alters the state of the particle, very careful measurements of the subsequent 
trajectories of both the particle and the rock can be used to accurately reconstruct the state of the particle before the collision, and
restore the particle to that state. But second law can put a limit on how much invasiveness is tolerable! If for example, the invasive
measurement involved setting fire and vaporising the particle, it would be practically impossible to regain the original state!

It is of course possible that the attributes change with time. The rose of the beginning could fade. Does this mean that the element of
reality that was so carefully constructed was not real at all? The physicist's answer to this is not to deny the element of reality or
its robustness, but to allow for a time evolution of states and their associated attributes. This is the idea of \emph{Dynamics}. The
determinism referred to earlier then takes the form of a \emph{Deterministic Dynamics}. These deterministic rules of dynamics not only ensure
unambiguous outcomes in ideal measurements, but they also ensure that no separate rules are necessary to describe measurements, unlike in
quantum theory.

\section{Ontology in Quantum Mechanics}
At least as per our present understanding, the standard quantum theory is \emph{inherently random}. From our previous discussion, no
strict ontology ought to be possible then. Quantum Measrements, as understood during the critical years of the development of quantum
theory, and as idealized by the \emph{Dirac-von Neumann} measurement models, are certainly invasive, and uncontrollably so. They are
invasive in an unpredictable way. This too leaves no room for an ontic description. The Born probability rule has to be invoked for a
consistent interpretation of quantum mechanics and that is where randomness becomes intrinsic.Paradoxically, the rules for time evolution
of states, or, quantum dynamics, is completely deterministic by itself! It is only measurement that brings in indeterminacy. Nevertheless,
as we shall see later, there are intriguing pointers to why there can be no ontic description of quantum mechanics, coming from purely
dynamical considerations(all unitary processes are considered dynamical here).

This also makes repeated measurements of the Dirac-von Neumann meaningless when performed on a single copy. The simple reason being
that the state after the first measurement bears no obvious relation to the state one started with, and the subsequent measurements 
can at best reveal the state after the first measurement. Therefore only \emph{ensemble} measurements become significant. For a good
account of the issues involved in getting information out of measurements on a single copy see \cite{alterbook}.

If there are such serious obstacles to ontology in quantum mechanics, why bother to go further? There are several good reasons for it!
Firstly, the extreme non-invasiveness of quantum measurements is certainly a distinctive feature of the Dirac-von Neumann, or more
precisely, \emph{Projective Measurements}. So the question naturally arises whether thre can be other measurement schemes that are
non-invasive or controllably non-invasive. It then becomes important to re-examine the ontology issue in the context of these alternate
measurement schemes. As it turns out, there are so many interesting alternatives to projective measurements today \cite{wisemannbook2010}.
It is the purpose of this article to do that examination carefully.

We set the following technical criterion for ontology: \emph{ontology is the ability to completely determine the previously unknown 
state of a single copy}. Even in cases where this is not possible, we introduce the notion of \emph{FAPP Ontology}(For All Practical Purposes)
as the ability to almost determine the unknown state of a single copy i.e state determination with specified amount of errors.

Though historically it was not recognized as such, we can now trace all the essential non-classical features of quantum theory to just
one principle, namely, \emph{The Principle of Superposition of States} \cite{diracbook1,diracbook2,bohr2013}. In fact, one can take
this principle to be the defining feature of quantum theories. Other aspects like \emph{Entanglement}, taken by many(particularly among
the Quantum Information community) to be the crux of quantum mechanics, is a natural consequence of the superposition principle.

It turns out that even without a very detailed analysis, one can show the impossibility of perfectly non-invasive measurements in
quantum mechanics by just invoking the superposition principle. We outline this powerful argument in Sec.(\ref{sec:superonto}).
Another, equally powerful argument against ontology can be given by invoking the \emph{No Cloning Theorem}. The proof of the No Cloning
Theorem involves only \emph{Unitarity}, and makes no reference to quantum measurements at all. It is surprising
that this theorem, which has nothing to do with the measurement process, could have such a strong bearing on the issue on ontology in
quantum mechanics. This second argument is presented in Sec.(\ref{sec:nocloneonto}). We then analyse the projective measurements
(Sec.(\ref{sec:projonto})), the protective measurements(Sec.(\ref{sec:protonto})), a method of cloning which we had named \emph{Information
Cloning}(Sec.(\ref{sec:infoonto})), the weak measurements(Sec.(\ref{sec:weakonto})), methods of approximate cloning (Sec.(\ref{sec:optimcloneonto})) for their implications on the question of ontology in quantum mechanics. 
\subsection{Superposition Principle and Ontology}
\label{sec:superonto}
Let us consider a hypothetical measurement device that is pefectly non-invasive i.e it leaves the system state undisturbed. We can consider
the initial unknown system state to be $|\psi\rangle_S$. Since this does not change, we can use a state-vector representation for the system. 
The treatment of the apparatus will be more subtle. All that the apparatus is required to do is produce a probability distribution of
outcomes which carries complete information about the expectation value of the observable in the system state $|\psi\rangle_S$. Therefore,
at least the final state of the apparatus ought to be described by a density matrix. Then one might as well describe the entire history
of the apparatus by a density matrix. Because the system stays in the same state throughout, it is consistent to treat the system by a
state vector, and the apparatus by a density matrix. The initial state of the system-apparatus composite can be taken to be:
\begin{equation}
\label{eq:SAinisuper}
|\psi\rangle_S\otimes\,\rho^A(0)
\end{equation}
Under the measurement ${\cal M}$, this goes to
\begin{equation}
\label{eq:meassuper}
|\psi\rangle_S\otimes\,\rho^A(0)\xrightarrow{{\cal M}}\,|\psi\rangle_S\otimes\rho^A(\langle\psi|O|\psi\rangle_S)
\end{equation}
The measurement ${\cal M}$ not being a \emph{Unitary} process, can take a pure density matrix to a mixed one. The final apparatus
(reduced)density matrix is in general mixed. If such a ${\cal M}$ could be realized, it can be used as often as necessary to measure
all the relevant observables for state tomography of $|\psi\rangle_S$ as the state is left undisturbed.

It is obvious that the map in eqn.(\ref{eq:meassuper}) does not preserve linear superpositions. More precisely, if the measurement
device works on $|\psi_1\rangle_S$ and $|\psi_2\rangle_S$, it will not work on a general superposition $\alpha|\psi_1\rangle+\beta|\psi_2\rangle$ i.e the measurement does not work on an \emph{arbitrary} unknown state.

This is a very powerful conclusion showing that the principle of linear superposition of states alone is enough to rule out ontology
in quantum mechanics and one need not invoke the deep, but confusing, chain of arguments invoked by the founders like Niels Bohr.
An explicit realization of this line of thinking is afforded by the measurements discussed in Sec.(\ref{sec:protonto}) and Sec.(\ref{sec:infoonto}). In the case of Protective Measurements, the scheme requires the unknown initial states to be \emph{non-degenerate} eigenstates
of a suitable Hamiltonian. A linear superposition of such states is no longer a state of the same type. In the case of Information Cloning, the scheme requires the unknown states to be Coherent States of a Harmonic Oscillator, and again, a superposition of such coherent states is not a
coherent state!
\subsection{The No-cloning theorem}
\label{sec:nocloneonto}
The No-Cloning theorem \cite{noclone} is one of the most striking of all results in quantum theory! Invoking nothing more than the inner-product
preserving nature of unitary transformations or the superposition principle, it states that no unitary process can ever
'copy' unknown quantum states. In a lighter vein it is said that there are no quantum Xerox machines! We shall first describe the theorem,
which is remarkably straightforward considering its profundity.

Consider an unknown state $|\psi\rangle_S$ of some quantum system and N identical copies of another, but \emph{known}, state
$|0\rangle_S$ of the same system(it is not really necessary that they be of the system, though). The latter are also called 'blanks'
or 'ancillaries'. A unitary transformation ${\cal U}$ acting on the \emph{tensor product} Hilbert space ${\cal H}^{N+1}$ is
said to be a \emph{universal cloning transformation} if it satisfies
\begin{equation}
\label{eq:univclone}
{\cal U}\,|\psi\rangle\otimes\,|0\rangle_1\otimes\ldots\otimes|0\rangle_{N}
=\,|\psi\rangle\otimes\,|\psi\rangle_1\otimes\ldots\otimes|\psi\rangle_{N}
\end{equation}
for every $|\psi\rangle$. The No-cloning theorem is a proof that no such universal unitary transformation can exist. For a proof
based only on unitarity of ${\cal U}$, consider a second state $|\chi\rangle$ so chosen that $|\langle \chi|\psi\rangle\|,\ne 0,1$.
Then the effect of ${\cal u}$ on $|\chi\rangle$ has to be
\begin{equation}
\label{eq:univclone2}
{\cal U}\,|\chi\rangle\otimes\,|0\rangle_1\otimes\ldots\otimes|0\rangle_{N}
=\,|\chi\rangle\otimes\,|\chi\rangle_1\otimes\ldots\otimes|\chi\rangle_{N}
\end{equation}
Taking the inner product between these two equations and using unitarity of ${\cal U}$, one gets,
\begin{equation}
\label{eq:cloneinner}
\langle \chi|\psi\rangle = (\langle \chi|\psi\rangle)^{N+1}
\end{equation}
But this is possible only if $|\langle \chi|\psi\rangle|=0,1$ which contradicts the initial premise about $|\chi\rangle$! The
same
proof can also be viewed as a consequence of the superposition principle.

What is the relevance of the No-cloning theorem to our discussion of ontology? The point is, that N can be made very very large, 
at least in principle, either in
a single application of the universal cloner or in many cascaded applications of it. Then we can set aside one out of N+1 copies
produced, and use the remaining N copies for an \emph{ensemble state detemination}. The accuracy of the subsequent state determination
can be improved with higher and higher N. One would still be left with one copy of the orginal unknown state even if the tomography
with the N copies is as invasive as can be. 

Thus if an universal cloner existed, one would in effect be able to make a non-invasive measurement on a single copy of an unknown state
and still be able to determine its state as accurately as one wishes. It is rather remarkable that this theorem which invokes only
aspects of unitary evolutions, with no explicit reference to quantum measurements, nevertheless captres the very essence of quantum
measures as per the Copenhagen Interpretation! This deep connection also borders on the mystic.

However, we shall introduce a novel \emph{Information Cloning} which bypasses the no cloning theorem in a subtle way and is a way
of getting information on a single copy, albeit with errors that can not be reduced arbitrarily.
\section{Projective Measurements and Ontology}
\label{sec:projonto}
Now we analyse why the Projective or Dirac-von Neumann measurements can not yield any ontology. Even though the arguments are
simple and straightforward, we recast them in the language of joint and conditional probabibilities so we can use the same framework
to address the issue of ontology in other contexts like weak measurements.

In a strong or projective measurement, the state after the first measurement is changed randomly to one of the eigenstates of
the observable being measured. 
The outcome of the apparatus is the corresponding eigenvalue. The fact that a given eigenstate-eigenvalue combination could have resulted from infinitely many unknown
initial states makes their reconstruction impossible from the information available after a single such measurement. 
Such a reconstruction requires an \emph{ensemble} measurements with optimally chosen 
observables. 

If repeated strong measurements are performed on a single copy, the second and all subsequent measurements are eigenstate measurements
where the eigenstate in question is the state after the first measurement. Therefore all subsequent measurements leave the system in
this same eigenstate and all subsequent apparatus outcomes are exactly the same as the outcome of the first measurement.
In other words, they do not generate any additional information required for the state reconstruction.
The strong measurements are not only highly invasive, they do not generate any information for determining the state.
These are the reasons, within standard quantum mechanics, for
the impossibility of an \emph{ontological} description.

Now let us recast these considerations in the language of conditional and joint probabilities of outcomes of repeated measurements 
on a single copy. Let the observable being measured is S, with the spectrum $s_i,|s_i\rangle_S$. If the initial unknown state
of the system is
\begin{equation}
\label{eq:projini}
|\psi\rangle_S =\sum\limits_i\,\alpha_i\,|s_i\rangle_S
\end{equation}
The probability distribution for the outcomes of the first measurement is given by
\begin{equation}
\label{eq:projprobfirst}
P(p_1) = \sum\limits_i\,|\alpha_i|^2\,\delta(p_1-s_i)
\end{equation}
This says that the first outcome is random with the above distribution. Let the outcome of the second measurement be $p_2$, and
as explained above, it has to be the same as $p_1$ because it is an eigenstate measurement. Therefore, the probability distribution
for $p_2$ is \emph{conditional} on the outcome $p_1$. In otherwords, the \emph{conditional probability distribution} $P(p_2|p_1)$
for the outcome $p_2$, conditional on the first outcome being $p_1$ is
\begin{equation}
\label{eq:projcondprob1}
P(p_2|p_1) = \delta(p_2-p_1)
\end{equation}
The \emph{Joint Probability Distribution} $P(p_2,p_1)$ for the outcomes of the first two of the repeated measurements is now
given by
\begin{equation}
\label{eq:projjoint2}
P(p_2,p_1) = P(p_2|p_1)\cdot P(p_1) = \sum\limits_i\,|\alpha_i|^2\,\delta(p_2-p_1)\delta(p_1-s_i) =
\sum\limits_i\,|\alpha_i|^2\,\delta(p_2-s_i)\delta(p_1-s_i)
\end{equation}
It is straightforward to generalize these to the outcomes of N repeated measurements on a single copy:
\begin{equation}
\label{eq:projjointN}
P(p_N,\ldots,p_1) = 
\sum\limits_i\,|\alpha_i|^2\,\prod\limits_{j=1}^N\,\delta(p_j-s_i)
\end{equation}
As usual, it is useful to introduce ${\mathrm y}_N$ to be the average of the first N outcomes, and consider its probability distribution
$P(y)$ i.e
\begin{equation}
\label{eq:genmeasmean}
{\mathrm y}_N = \frac{\sum_i\,p_i}{N}
\end{equation}
and
\begin{equation}
\label{eq:meandist}
P({\mathrm y}_N) =\int\ldots\int\,\prod\limits_i\,dp_i\,P(\{p\})\,\delta({\mathrm y}_N - \frac{\sum_i\,p_i}{N})
\end{equation}
On using eqns.(\ref{eq:projjointN},\ref{eq:meandist}), it follows that
\begin{equation}
\label{eq:projmeandist}
P({\mathrm y}_N) = \sum\limits_i\,|\alpha_i|^2\,\delta({\mathrm y}_N-s_i)
\end{equation}
The repeated measurements have not changed the nature of the distribution at all, and it remains the same as eqn.(\ref{eq:projprobfirst})!
Though our simple reasoning had already told us this, the formalism of conditional and joint probabilities used above will prove to be
useful in more complicated situations where there are no such simple reasonings available.
\subsection{Sharpening the ontology criterion}
The form of the eqn.(\ref{eq:projmeandist}), derived for Projective Measurements which are decidedly invasive and hence incapable of any
ontological descriptions, suggests an even more precise technical criterion for ontology. For that, let us contrast eqn.(\ref{eq:projmeandist})
with what one would expect in the case of ensemble measurements on the basis of the Central Limit Theorem:
\begin{equation}
\label{eq:meandistclt}
P({\mathrm y}_N) = N\,e^{-\frac{N\,({\mathrm y}_N-\mu)^2}{\Delta^2}}
\end{equation}
This suggests a way to sharpen the criterion for onticity in quantum mechanics, given verbally earlier, to the following precise mathematical 
criterion: \emph{exact} ontology
in quantum mechanics is the ability to find non-invasive measurement schemes such that the mean of the N outcomes of repeated measurements on a \emph{single}
copy of a system in an unknown state takes the deterministic form
\begin{equation}
\label{eq:exactonto}
P({\mathrm y}_N) = \delta ({\mathrm y}_N-\mu)\quad\quad \mu = \langle\psi|O|\psi\rangle
\end{equation}
Not surprisingly, there will be no candidates within quantum mechanics for this criterion.

The next best possibility will be the \emph{FAPP-Ontology} discussed earlier. The following two criteria provide precise characterizations
of such. The first is that the statistics of outcomes of repeated measurements on a single copy will be very similar to that obtained
from measurements on an ensemble. In particular, the distribution for the average ${\mathrm y}_N$ will be a \emph{single} distribution
as in eqn.(\ref{eq:meandistclt}), and additionally $\mu = \langle\psi|S|\psi\rangle$. The figures of merit for the FAPP ontology are
i) how close $\mu$ actually is to the expectation value, and ii) how small the error $\epsilon = \frac{\Delta}{\sqrt{N}}$ is.

The second criterion allows for the distribution $P({\mathrm y}_N)$ to deviate from a single distribution but with very small deviations
i.e
\begin{equation}
\label{eq:fapponto2}
P({\mathrm y}_N) = p_0\,e^{-\frac{(y_N-\langle S \rangle_\psi)^2}{\epsilon^2}}+\sum\limits_i\,p_i\,e^{-\frac{({\mathrm y}_N-\mu_i)^2}{\epsilon_i^2}}
\end{equation}
In this case, the avearage outcome of repeated measurements will be \emph{random}, and ensemble measurements become a necessity; measurements
on a single copy will not reveal \emph{any} information about the unknown state. In the coming sections we shall discuss explicit realizations of these criteria. 
\section{Protective Measurements and Ontology}
\label{sec:protonto}
Aharonov, Anandan and Vaidman \cite{protect} proposed a remarkable type of experiments which they called \emph{Protective Measurements}. They gave an
explicit realization for them and showed that for a \emph{restricted} class of states, and in a certain \emph{ideal} limit, one could get 
full information about single copies of such restricted class of states \emph{without} affecting the state. From whatever we have said so far, such a
proposal would realize exact ontology in the ideal limit. Closer examination, however, shows that even these remarkable category 
of measurements actually provide only FAPP ontology, as the ideal limit requires measurements lasting infinitely long. Now we elaborate 
on the details.

They consider states that are \emph{non-degenerate} eigenstates of some \emph{unknown} Hamiltonian. For this reason, the states are
indeed unknown.
   Let us briefly review the standard projective measurements to see the differences and commonalities between projective and protective
measurements. For every type of measurement it is necessary to characterize the measuring apparatus. Niels Bohr was of the opinion that this
necessarily had to be \emph{classical}, whereas Dirac and von Neumann found it desirable to take this also to be a quantum system. It is also 
important to consider the modern picture of the Dirac-von Neumann Scheme. According to this, the final act of the measurement(the
one that breaks the so called \emph{infinite von Neumann regression}) is 
\emph{environmental decoherence} which accounts for the real life situation that there is a complex environment with which both the system
and the apparatus are interacting. This, technically speaking, renders the final density matrix diagonal in an apparatus Hilbert space
basis which defines the \emph{Pointer States} for the apparatus. Let $R_A$ be the observable of the apparatus whose eigenstates are the pointer
states.
In the Dirac-von Neumann measurement theory formalism, one introduces an apparatus operator $Q_A$ that is canonically conjugate to $R_A$ i.e $[R_A,Q_A] = i\hbar$

For both types of measurements, the interaction between the 'apparatus' and the system is taken to be described by a
Hamiltonian:
\begin{equation}
\label{eq:measint}
H_I(t) = g(t)Q_A\,S \quad\quad \int\,g(t)dt=1
\end{equation}
Here S is the system observable that is being measured and $Q_A$ the observable of the apparatus described above. The integral condition on g(t) is a convenient normalisation which can be taken without loss of generality. In addition to this
interaction Hamiltonian, the time evolution of both the system and the apparatus are respectively governed by their own Hamiltonians
$H_A$ and $H_S$, respectively.

The projective measurements correspond to an \emph{impulsive} $g(t)$ i.e
   $g(t)$ is non-zero only in a very small time interval $-\frac{\epsilon}{2} < t < \frac{\epsilon}{2}$.
   The time-evolution unitary transformation taking pre-measurement-interaction states to post-measurement-interaction states is
given by
\begin{equation}
\label{eq:measunitary}
U(\frac{\epsilon}{2},-\frac{\epsilon}{2}) = e^{-\frac{i}{\hbar}~\int_{-\frac{\epsilon}{2}}^{\frac{\epsilon}{2}}~ H~dt}
\end{equation}
Normally this unitary transformation is given by the so called \emph{time ordered integral} over the \emph{total} Hamiltonian
H(t):
\begin{equation}
\label{eq:totalham}
H(t) = H_A+H_S+H_I
\end{equation}
   In the limit of the measurement interaction being extremely impulsive i.e $\epsilon \rightarrow 0$, the time ordered integral is
well approximated by
\begin{equation}
\label{eq:impulsiveunitary}
U = e^{-\frac{i}{\hbar}~Q_A\,S}
\end{equation}
It should be noted that $H_A, H_S$ do not contribute in this impulsive limit(it is understood that these Hamiltonians are bounded).
   The combined state of the system and apparatus before measurement is taken to be the disentangled state
\begin{equation}
\label{eq:measini}
|t_<\rangle = |\nu\rangle_S|\Phi(r_0)\rangle_A
\end{equation}
The initial apparatus state, in the Dirac-von Neumann scheme is taken to be an eigenstate of $R_A$ with an eigenvalue, say, $r_0$;
this corresponds to the initial reading of the apparatus. 
To avoid technical difficulties arising out of the use of continuous variables, the initial apparatus state $|\Phi(r_0)\rangle$ is taken to be
a wavepacket \emph{sharply} centred around the
value $r_0$ of $R_A$. 

   Here $|\nu\rangle$ is the unknown system state on which a measurement of the observable $S$ is
performed. 
   If $|s_i\rangle$ are the eigenstates of $S$ i.e $S|s_i\rangle = s_i|s_i\rangle$, and $|\nu\rangle=\sum_i \alpha_i|s_i\rangle$ the post-measurement interaction state is given by
\begin{equation}
\label{eq:postprojstate}
|t_>\rangle = U|t_<\rangle =\sum_i~\alpha_i~e^{-\frac{i}{\hbar}s_i~Q_A}~|s_i\rangle~|\Phi(r_0)\rangle
\end{equation}
   As $Q_A$ is canonically conjugate to $R_A$, the exponential operator shifts the value of $R_A$ by $s_i$
and one gets the \emph{entangled} state
\begin{equation}
\label{eq:postprojstate2}
|t_>\rangle = \sum_i~\alpha_i~|s_i\rangle~|\Phi(r_0+s_i)\rangle
\end{equation}
This explicitly manifests the one-one correspondence between the states $|s_i\rangle$ of the system, and the states $|\Phi(r_0+s_i)\rangle$
of the apparatus. But the state in eqn.(\ref{eq:postprojstate2}) is \emph{entangled} and it hardly reflects the single outcomes expected of
a good measurement! It is instructive to see how decoherence 'solves' this issue; for that, consider the \emph{pure} density matrix corresponding
to this state:
\begin{equation}
\label{eq:postprojrho}
\rho^{S+A}(t_>) = \sum\limits_{i,j}\,\alpha_i\,\alpha_j^*\,|s_i\rangle\langle s_j|\,|\Phi(r_0+s_i)\rangle\langle \Phi(r_0+s_j)|
\end{equation}
Clearly this matrix is not diagonal in the pointer basis $|\Phi\rangle$. Decoherence reduces this to the mixed density matrix
\begin{equation}
\label{eq:postprojrhomixed}
\rho^{S+A}(t_>) = \sum\limits_{i}\,|\alpha_i|^2\,|s_i\rangle\langle s_i|\,|\Phi(r_0+s_i)\rangle\langle \Phi(r_0+s_i)|
\end{equation}
Though this still does not explain how single outcomes come about, it has at least reduced that to a classical problem of picking
from a mixture, much like picking a card out of a deck.

To pictorially contrast the projective and protective cases, we schematically show in the next figure the outcomes of a standard 
Stern-Gerlach experiment viewed as a projective measurement.
\begin{figure}[htp!]
  \centering
  \includegraphics[width=2.5in]{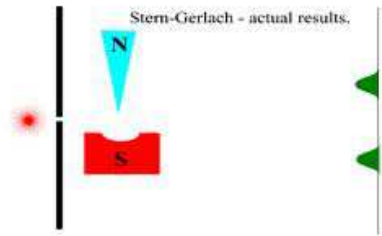}
  \caption{The Stern-Gerlach Measurement}
  \label{sterngerlach}
\end{figure}

With this background, it is easy to grasp the essentials of a \emph{Protective Measurement}.
The major departure from projective measurements is that now the measurement interaction behaves oppositely to what it did
in the case of projective measurements i.e
   the interaction time $T$ is now taken to be very large, approaching infinity! Let us leave aside for now questions like the meaning
of measurements that take infinitely long, and proceed.  It is simplest to take $g(t)$ to be a constant. Then the normalization
condition gives $g =\frac{1}{T}$, where T is the long duration of the measurement, which will tend to $\infty$ in the end.
   The total Hamiltonian becomes
\begin{equation}
\label{eq:tothamprot}
H = H_A + H_S + \frac{1}{T}~Q_A~S
\end{equation}
which is \emph{time independent}. Again, for simplicity we restrict analysis to the choice $[H_A,Q_A]=0$(for a complete
discussion of the general situation see \cite{unprotect}). However, even in the standard Stern-Gerlach case, such a simplification
does not happen. This condition allows both $H_A, Q_A$ to be simultaneously diagonalized i.e we can take
\begin{equation}
\label{eq:simevprot}
Q_A|a_i\rangle_A = a_i|a_i\rangle_A\quad\quad H_A|a_i\rangle_A = E^A_i\,|a_i\rangle_A
\end{equation}

   $H_S$ taken to be \emph{unknown} has the non-degenerate eigenstates $|j\rangle_S$, with eigenvalues $\omega_j$. Because of the
simplifying assumptions made, the total Hamiltonian $H$ also commutes with $Q_A$ and both of them can also be diagonalized
simultaneously.
   If we take $H_A|a_i\rangle_A = E^A_i|a_i\rangle_A$, the simultaneous eigenstates of $H$ and $Q_A$ are of the form 
$|j,i\rangle_S\,|a_i\rangle_A$ with $|j,i\rangle_S$ satisfying
\begin{equation}
\label{eq:protsysham}
(H_S+\frac{1}{T}~a_i~S)|j,i\rangle_S = \Omega(j,i)|j,i\rangle_S
\end{equation}
It is clear that
\begin{equation}
\label{eq:protsysev}
\Omega(j,i)\xrightarrow{T\rightarrow\infty}\:\omega_j\quad\quad |j,i\rangle_S \xrightarrow{T\rightarrow\infty}\:|j\rangle_S
\end{equation}
The eigenvalues and eigenstates of the total Hamiltonian $H$ can now be expressed as
\begin{equation}
\label{eq:prottotH}
H|j,i\rangle_S\,|a_i\rangle_A = E(j,i)|j,i\rangle_S|a_i\rangle_A=(E^A_i+\Omega(j,i))|j,i\rangle_S|a_i\rangle_A
\end{equation}
For very large T, $\Omega(j,i)$ can be calculated in first order perturbation theory to get 
\begin{equation}
\label{eq:protpert}
\Omega(j,i)=\omega_j+\frac{1}{T}\cdot a_i\,\langle j|S|j \rangle_S 
\end{equation}
If the unknown system state before measurement is the non-degenerate eigenstate $|k\rangle_S$ of $H_S$, the joint state 
before measurement is taken to be $|k\rangle_S~|\Phi(r_0)\rangle_A$, with $|\Phi(r_0)\rangle_A$ being the same as what
was used in projective measurements.
 
   The joint state after time $T$ is 
\begin{equation}
\label{eq:protpost}
|k,T\rangle = U(T)|k\rangle_S|\Phi(r_0)\rangle_A =\sum\limits_{i,j}\langle a_i|\Phi(r_0)\rangle_A~\langle j,i|k\rangle_S
e^{-\frac{i}{\hbar}TE(j,i)}~|j,i\rangle_S|a_i\rangle_A
\end{equation}
   In first order perturbation theory, $\langle j,i|k\rangle_s = \delta_{j,k}$. Putting everything together
\begin{equation}
\label{eq:protpostpert}
|k,T\rangle~\xrightarrow{T\rightarrow\infty}~~e^{-\frac{i}{\hbar}~\omega_k~T}|k\rangle_S\cdot e^{-\frac{i}{\hbar}H_A~T}\cdot e^{-\frac{i}{\hbar}\langle k|S|k \rangle_S~Q_A}~|\Phi(r_0)\rangle_A
\end{equation}
   In other words
\begin{equation}
\label{eq:protpostpert2}
|k,T\rangle~\rightarrow~~e^{-\frac{i}{\hbar}~\omega_k~T}|+\rangle\cdot e^{-\frac{i}{\hbar}H_A~T}\cdot 
~|\Phi(r_0+\langle k| ~S| k\rangle)\rangle_A
\end{equation}
Thus under these protective measurements,   \emph{the original state is protected and the apparatus reads the expectation value 
$\langle k|S|k \rangle_S$}!
This is modulo the $e^{-\frac{i}{\hbar}H_A~T}$ factor. 

Since the state is 'undisturbed', one can reuse it for carrying out protective measurements
of all the necessary observables for complete state determination.
 The apparatus and the system are \emph{disentangled}, and there is no need to take recourse to decoherence to achieve the final step
in the measurement process.
   This is what can be called the \emph{Ideal} protective measurements, in the sense that it is valid only in the strict $T=\infty$ limit.
The next figure shows the situation to be expected for an ideal protective Stern-Gerlach expt.
\begin{figure}[htp!]
  \centering
  \includegraphics[width=2.5in]{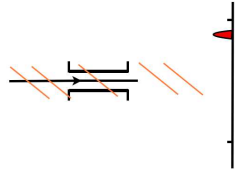}
  \caption{The Ideal Protective Measurement}
  \label{idealprotective}
\end{figure}
Unlike the standard Stern-Gerlach set up, the silver atoms in an ideal protective measurement would strike the screen at only
\emph{one} spot, in between the extreme positions encountered in the standard case. Its location is a precise measure of the \emph{expectation
value} of the measured observable in the unknown initial state.

But eqn.(\ref{eq:protpostpert2}) is precisely the kind that had been argued to be in conflict with the superposition principle in 
Sec.(\ref{sec:superonto})! But the AAV scheme cleverly evades this by considering the unknown initial states to be non-degenerate
eigenstates of $H_S$; therefore, superpositions of such states can no longer be non-degenerate eigenstates of $H_S$!
\subsection{Non ideal protective measurements}
\label{sec:nonidealprot}
   The Ideal case is obviously unphysical as it is meaningless for any measurement to last infinitely long! 
In real life situations $T$ can be very very large(compared to the time scales involved) but not $\infty$.
One may naively argue that for all practical purposes the difference between such very large T and the ideal limit should be
negligible.
Indeed, for ensemble measurements the difference between very large $T$ and $T=\infty$ is negligible in the precise sense that the
resulting probability distributions for outcomes differ only very slightly, and all the statistical conclusions are not affected
significantly.

   But for measurements on single copies, which are the only relevant measurements in the context of ontology, the situation is
\emph{dramatically} different. In Quantum Mechanics, unlike in the classical counterpart, individual outcomes of measurements
are completely random and unpredictable. Even outcomes with hopelessly small probabilities can manifest. Only if their probability is
\emph{exactly} zero, will they not show up. This makes a very significant difference for protective measurements. In a nutshell, 
departure from $T=\infty$ causes a very small but significant entanglement between the system and the apparatus. This can cause the first
protective measurement to project the unknown initial state into any state that is orthogonal to it. This way, not only is the state
not protected during the first measurement, it renders meaningless the outcome of even protective measurements subsequently. no state
reconstruction is possible and there is no strict ontology. This was the criticism of protective ontology that was made by both \cite{unprotect}
as well as by \cite{alter}.
 
   To address these issues we need to consider all sources of $\frac{1}{T}$ corrections to the ideal results.
   We refer the reader to \cite{unprotect,anirban,75yrs} for the technical details. Here  we shall list the important sources of $\frac{1}{T}$ corrections
and discuss their importance.
   In the sum of eqn.(\ref{eq:protpost}), one will have to take into account system states $|j\ne k,i\rangle_S$.
In order to get the leading $\frac{1}{T}$ corrections, \emph{second order} perturbation theory becomes necessary.
   This typically introduces corrections of the type $|k^\prime\rangle_S\cdot Q_A^2|\Phi(r_0)\rangle$.
   Schematically the effect of these corrections can be represented as
\begin{equation}
\label{eq:protpertcorr}
|T\rangle = |ideal\rangle +\frac{c}{T}|Non-ideal\rangle
\end{equation}
   It is important to note that $\langle ideal|Non-ideal\rangle = 0$ because of the nature of perturbation theory. Now we can further
enumerate some possibilities:
\begin{itemize}
   \item State is protected and apparatus reads the $\langle Q_S\rangle$ in that state with $P=1-\frac{c^2}{T^2}$.
   \item State protected but pointer in \emph{all possible} states with probability $\simeq\frac{1}{T^2}$.
   \item State collapses to the state \emph{orthogonal} to it and the pointer reads the expectation value in the
orthogonal state with probability $\simeq\frac{1}{T^2}$.
   \item State collapses to the orthogonal but pointer in all possible states with probability $\simeq\frac{1}{T^2}$
\end{itemize}
   This is depicted pictorially in the next figure.
It is worth emphasizing that in each of these cases, the system state after measurements remains correlated with the original state. This is in sharp contrast to projective measurement where the system state after the measurement is completed has no memory of the original state whatsoever.
\begin{figure}[htp!]
  \centering
  \includegraphics[width=2.5in]{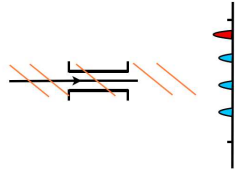}
  \caption{The Non-ideal Protective Measurement}
  \label{nonidealprotective}
\end{figure}
\subsection{Adiabatic two qubit interactions -- another twist}
As a further generalization of the protective measurement schema, Anirban Das and myself \cite{anirban} considered the case where the role of the
apparatus is also played by another qubit or by a quantum system with finite dimensional Hilbert space.   
Let us illustrate with the example of the qubit as a detector. We take the basis states to be $|d_{\uparrow}\rangle_A$ and $|d_{\downarrow}\rangle_A$. The system is also taken to be a qubit with its Hilbert space spanned by $|\uparrow\rangle_S$ and $|\downarrow\rangle_S$.
   The measurement interaction is taken to be represented by the unitary transformation ${\cal U}$: 
\begin{equation}
\label{eq:twoqubitU}
|\uparrow\rangle|d_{\downarrow}\rangle\,\xrightarrow{{\cal U}}\,|\uparrow\rangle\,|d_{\uparrow}\rangle\quad\quad
|\downarrow\rangle|d_{\downarrow}\rangle\,\xrightarrow{{\cal U}}\,|\downarrow\rangle\,|d_{\downarrow}\rangle
\end{equation}
   The components of spin are taken to be along the z-axis for both systems.
   For projective measurements where there are only two possible outcomes, it suffices
to take $|d_{\uparrow}\rangle,|d_{\downarrow}\rangle$ as the pointer states. Whether there are any realistic 'environments' that
can result in decoherence in this basis is not very clear.
   For adiabatic measurements where there can be a near-continuum of outcomes, we shall take
\emph{angular momentum coherent states} obtained by rotating, say, $|d_{\uparrow}\rangle$ by $\theta$ about
the x-axis as the pointer states. Once again the existence of suitable decoherence mechanisms in this basis remains to be understood.
   More general possibilities for pointer states can also be considered.
   The interaction Hamiltonian that generates the unitary transformation ${\cal U}$ turns out to be(actually there are infinitely many such Hamiltonians!)
\begin{equation}
\label{eq:2qubitham}
-\pi~g(t)~P^S_{z,+}\otimes~P^A_{x,-}
\end{equation}
   Here $P_{a,\pm}$ are the \emph{projection operators} for spin $\pm$ along the $a$-direction.
   $H_A$ is taken to be the rotationally invariant ${\vec S}_A\cdot{\vec S}_A$. This, being a constant, does not lead to any pointer state broadening.
   Let the initial unknown system state be
\begin{equation}
\label{eq:2qubitini}
|\nu\rangle = \alpha|\uparrow\rangle_S + \beta|\downarrow\rangle_S
\end{equation}
In the ideal limit, protective measurements of this type maintain the original state and the pointer state is $\theta = \pi|\alpha|^2$.
But here too the non-ideal case is the morerealistic and we enumerate the possible outcomes \cite{anirban,75yrs}.
\begin{itemize}
\item   After accounting for the relevant $\frac{1}{T}$ corrections also, the dominant outcome is when the original state is protected
and the apparatus outcome is the expectation value of the observable in the original state. But unlike the ideal case, the probability
of this happening is no longer unity; instead it happens with probability $P\simeq 1-\frac{c^2}{T^2}$.
\item   State collapses to its orthogonal; unlike the protective measurements considered so far, the apparatus state now is 
uniquely determined to be
\emph{$|d_{\downarrow}\rangle_x$}! This happens
 with probability $P\simeq\frac{1}{T^2}$.
\item   State is protected but apparatus again in
\emph{$|d_{\downarrow}\rangle_x$}
, with probability $P\simeq\frac{1}{T^2}$.
\item   It is intriguing that the 'failed' protective measurements now always produce the same apparatus state
\emph{$|d_{\downarrow}\rangle_x$} .
\end{itemize}
We see that because of nonvanishing probabilities for deviations from the ideal case, perfect ontology is not possible. The last
point mentioned above i.ethe failed cases coming with a well defined apparatus state might give hope that the lack of perfect
ontology may somehow be overcome by exploiting this feature. Even though the apparatus state, being fixed, does not convey any 
information about the initial system state, the state of the system after the measurement being just orthogonal to it, carries
all information about it. Unfortunately, no universal unitary transformation can transform an unknown initial state of a qubit
to its orthogonal state. But what is worse, there is no way to tell, with only single copies, that the protective measurement
has actually failed. The reason is that
as long as the pointer states are the ones produced by rotating $|d_{\downarrow}\rangle$ through
$\theta$ around the x-axis, 
\emph{$|d_{\downarrow}\rangle_x$} will have to be expressed as $\frac{1}{\sqrt{2}}(|d_{\uparrow}\rangle
+|d_{\downarrow}\rangle$. This being a superposition of pointer states, there are finite probabilities for different outcomes,
and the failed case will behave as in a projective case.
   This again precludes any perfect ontological significance to these unknown states.

   However, protection fails with very low probability.
   This means protective measurements can give practically full information about a class of unknown states in such a
way as to protect the \emph{purity} of the post-measurement ensemble to a very high degree.
   Further, a dramativc decrease in the \emph{size} of the ensemble for state tomography is possible.
In other words, protective measurements can provide FAPP ontology to an arrbitrary degree, and this
can be important and highly useful in this pragmatic sense \cite{pracprot,75yrs} though from a
philosophical point of view they can not deliver the ontological goods.
Because of all these interesting aspects, it is critical that they are subjected to a proper experimental study. For some feasible
suggestions, the reader is referred to \cite{75yrs}.
\section{Weak Measurements and Ontology}
\label{sec:weakonto}
Now we take up another class of remarkable measurement schemes called \emph{Weak Measurements} and \emph{Weak Value
Measurements}. These were also discovered by Aharonov, and his collaborators \cite{AVA;PRL1988,AV;PRA1989} . Let
us first dispose off the weak value measurements as they are by design unsuited for ontology. These are also called measurements
with \emph{Post-selection}; a post-selection of the system state is made through a projective measurement, following a weak
measurement on an initial, possibly unknown state. Obviously, the projective measurements involved in the post-selection stage
are invasive on the system. For this reason, this class of measurements can not have any bearing on the issues of ontology discussed here.

\subsection{Weak Measurements Without Post-Selection}
On the other hand, if no post-selection is made, removing thereby the invasive elements, weak measurements on their own appear to
be ideal for the ontological issues. As their name suggests, they are \emph{minimally invasive}, with this degree of invasiveness
apparently under full control. Here too, it is possible to make such measurements both on ensembles and on single copies. We consider
only the latter here.

As in Sec.(\ref{sec:projonto}), let S be the observable of the system with $s_i,|s_i\rangle_S$ its spectrum, which we take to be
\emph{non-degenerate}. 
The initial states of the system and the apparatus are taken to be \emph{pure} and as in eqn.(\ref{eq:measini}).
The measurement interactions are also of the form of eqn.(\ref{eq:measint})
discussed in Sec.(\ref{sec:protonto}). But there is an important difference now in that $Q_A$ need not be as restrictive
as in the Dirac-von Neumann measurement schemes.

The \emph{Pointer States} of the apparatus denoted by $|p\rangle_A$, are taken to be eigenstates of
an apparatus observable $P_A$. The point of view taken here is that such pointer states form the basis in which the density matrix 
becomes diagonal as a result 
of \emph{decoherence}. They are not always labelled by the mean values of $P_A$ in a given state of the apparatus.
Therefore, the specification of an \emph{apparatus} involves some quantum system, along with a decoherence mechanism which picks out
the pointer states. 
The $P_A, Q_A$ pair need not be canonically conjugate.
A detailed account of many important aspects of weak measurements can be found
in \cite{kofman2012}. 
In what follows we shall nevertheless stick to the canonical pair for convenience.

The initial aparatus states are
taken to be \emph{Gaussian states} centred around some $p_0$. In other words, for $p_0=0$,
\begin{equation}
|\phi_0\rangle_A = N\,\int\,dp\,e^{-\frac{p^2}{2\Delta_p^2}}\,|p\rangle_A\quad\quad N^2\sqrt{\pi\Delta_p^2}=1
\label{eq:iniapp}
\end{equation} 
In projective measurements, the Gaussians are taken to be very narrow i.e $\Delta_p\,<<\,0$ so that they approximate pointer states to a high
degree. In contrast, for weak measurements, $\Delta_p >> 1$.That means that the initial apparatus state is a \emph{very broad} superposition of pointer states with practically \emph{equal} weight for
many pointer states. Though even in the weak case, the initial apparatus state is also peaked at $p_0=0$, it is \emph{not} a pointer state.
This important point has led to confusing statements in literature. 

The measurement interaction is still taken to be \emph{impulsive} i.e the function g(t) is
nonvanishing only during a very small duration, say, $-\epsilon < t < \epsilon$. 
We leave out the details (the reader is referred to \cite{ndhweak} for them) and give only the essential results.
The \emph{post-measurement} density matrix turns out to be:
\begin{equation}
\rho^{post}_{SA} = \int dp\, |N(p,\{\alpha\})|^2|p\rangle\langle p|_A\,|\psi(p,\{\alpha\}\rangle\langle \psi(p,\{\alpha\}|_S
\label{eq:weakpost}
\end{equation}
where
\begin{equation}
N(p,\{\alpha\}) = N\,\sqrt{\sum_i\,|\alpha_i|^2\,e^{-\frac{(p- s_i)^2}{\Delta_p^2}}}
\label{eq:Nweakpost}
\end{equation}
\begin{equation}
|\psi(p,\{\alpha\}) = \frac{N}{N(p,\{\alpha\})}\,\sum_j\,\alpha_j\,e^{-\frac{(p- s_j)^2}{2\Delta_p^2}}\,|s_j\rangle_S
\label{eq:psiweakpost}
\end{equation}
For an \emph{ensemble} of weak measurements, $P(p,\{\alpha\})=|N(p,\{\alpha\})|^2$ being the probability for outcome p, the mean outcome is
\begin{equation}
\langle p \rangle_\psi = \int dp\, p|N(p,\{\alpha\})|^2 = \sum_i\,|\alpha_i|^2\,s_i 
\label{eq:weakmean}
\end{equation}
yielding the same expectation value as in strong measurements. 
The variance of the outcomes can be readily calculated to yield
\begin{equation}
(\Delta p)^2_\psi = (\Delta p)^2+(\Delta S)^2
\label{eq:weakvariance}
\end{equation}
This exposes one of the major weaknesses(!) of weak measurements i.e the errors in individual measurements are huge. This can be 
reduced statistically as usual. If one considers averages over $M_w$ measurements, the variance in the average, 
is $\frac{\Delta_p}{\sqrt{2M_w}}$. It makes sense to compare different measurement schemes only for a \emph{fixed} statistical error.
Therefore if averaging is done over $M_s$ strong measurements, 
\begin{equation}
\frac{\Delta S}{\sqrt{M_s}}=\frac{\Delta_p}{\sqrt{2M_w}}\rightarrow M_w = (\frac{\Delta_p}{\Delta S})^2\,\frac{M_s}{2}
\label{eq:sizeweakmeas}
\end{equation} 
The required resources will be supermassive! 

The aspect of weak measurements that has gained great prominence is its alleged \emph{non-invasiveness}. 
One possible measure of this non-invasiveness is provided by the \emph{post-measurement reduced density matrix} of the system:
\begin{equation}
\rho^{post}_S = \rho^{ini} - \frac{1}{4\Delta_p^2}\,\sum_{i,j}\,(s_i-s_j)^2\,\alpha_i\alpha_j^*\,|s_i\rangle \langle s_j|
\label{eq:weakpostreduced}
\end{equation}
Thus, for very large $\Delta_p$, the reduced density matrix of the system practically equals that of the initial state.

The combination of an \emph{exact} estimate for the expectation value, as given by eqn.(\ref{eq:weakmean}), as well as 
the  maintenance of the state to a very high degree as per eqn.(\ref{eq:weakpostreduced}) may give rise to the expectation that 
weak measurements may offer the best hopes for ontology in quantum mechanics. What would make such an expectation particularly
exciting is that these measurements can be done on \emph{any state} i.e they appear to offer FAPP ontology for arbitrary states!
We investigate this by turning to an analysis of 
\emph{repeated weak measurements on a single copy} as given in \cite{ndhweak} with particular emphasis on ontology. Two aspects that need to be
particularly focussed upon in this context are i) how closely the averages of N outcomes approximate the exact expectation values, and, ii)
how the single state gets degraded as a result of multiple weak measurements.

The following schema defines for us repeated weak measurements of the same observable on a single copy \cite{ndhweak}: (i) perform a weak measurement of system observable S
in state $|\psi\rangle_S$
with the apparatus in the state of eqn.(\ref{eq:iniapp}) with \emph{very large} $\Delta_p$, 
, ii) let the definitive outcome, defined as above, be $p_1$, and the single system state be $|\psi(p_1,\{\alpha\})\rangle_S$, iii) restore 
the apparatus to its initial state, and, 
iv) repeat step (i), and so on. After N such steps, let the sequence of outcomes
be denoted by $p_1,p_2\ldots,p_N$ and the resulting system state by $|\psi(\{p\},\{\alpha\})\rangle_S$.

The probability distribution for the first outcome $p_1$,$P^{(1)}(p_1)$ is given by 
\begin{equation}
\label{eq:weakrepeatfirst}
N^{(1)}(p_1,\{\alpha\})|^2=|N(p_1,\{\alpha\})|^2
\end{equation} 
with $N(p,\{\alpha\})$
given by eqn.(\ref{eq:Nweakpost}). The corresponding system state is given by $|\psi(p_1,\{\alpha\})\rangle_S$ of eqn.(\ref{eq:psiweakpost}). 
Thus the set of $\alpha$
for this state is given by
\begin{equation}
\alpha_i^{(1)} = \frac{N}{N(p_1,\{\alpha\})}\,e^{-\frac{(p_1-s_i)^2}{2\Delta_p^2}}\,\alpha_i
\label{eq:psiweakfirst}
\end{equation}
Since in step (iii) the apparatus state has been restored, the probability distribution $P^{(2)}(p_2)$ for the outcome $p_2$ at the end of
the second weak measurement, is given by 
\begin{equation}
P^{(2)}(p_2) = 
|N^{(2)}(p_2,\{\alpha\})|^2=|N^{(1)}(p_2,\{\alpha^{(1)}\})|^2 
\end{equation}
Substituting from eqn.(\ref{eq:psiweakfirst}), one gets
\begin{equation}
P^{(2)}(p_2) = \frac{(N^2)^2}{P^{(1)}(p_1)}\,\sum_i\,|\alpha_i|^2\prod\limits_{j=1}^2e^{-\frac{(p_j-s_i)^2}{\Delta_p^2}}
\label{eq:weakprob2}
\end{equation}
As stressed in \cite{ndhweak}, $P^{(2)}(p_2)$ is actually the \emph{conditional probability} $P(p_2|p_1)$ of obtaining $p_2$
conditional to having already obtained $p_1$ (that is the reason for the explicit dependence on $p_1$ in eqn.(\ref{eq:weakprob2})).
The \emph{joint probability} distribution $P(p_1,p_2)$ is therefore given by $P(p_2,p_1) = P(p_2|p_1)P(p_1)$ to give
\begin{equation}
P(p_1,p_2) = (N^2)^2\,\sum_i|\alpha_i|^2\,\prod\limits_{j=1}^2\,e^{-\frac{(p_j-s_i)^2}{\Delta^2}}
\label{eq:probweak2}
\end{equation}
The state after the second measurement is given by the exact analog of eqn.(\ref{eq:psiweakfirst}):
\begin{equation}
\alpha_i^{(2)} = \frac{N}{N^{(2)}(p_2,\{\alpha^{(1)}\})}\,e^{-\frac{(p_2-s_i)^2}{2\Delta_p^2}}\,\alpha_i^{(1)}
\label{eq:psiweaksec}
\end{equation}
It is useful to explicitly write this state:
\begin{equation}
|\psi(p_1,p_2,\{\alpha\}) = \frac{\sum\limits_i\,\prod\limits_{j=1}^2\,e^{-\frac{(p_j-s_i)^2
}{2\Delta_p^2}}\,\alpha_i|s_i\rangle_S}{\sqrt{\sum\limits_i\,|\alpha_i|^2\prod\limits_{j=1}^2 e^{-\frac{(p_j-s_i)^2}{\Delta_p^2}}}}
\label{eq:psiweak2}
\end{equation}
It is remarkable that these results are all symmetric in the outcomes $p_i$.
Eqns.(\ref{eq:probweak2},\ref{eq:psiweaksec}) readily generalize to the case of M repeated measurements:
\begin{equation}
P(p_1,\ldots,p_M) = (N^2)^M\,\sum_i|\alpha_i|^2\,\prod\limits_{j=1}^M\,e^{-\frac{(p_j-s_i)^2}{\Delta^2}}
\label{eq:probweakN}
\end{equation}
\begin{equation}
|\psi(p_1,\ldots,p_M,\{\alpha\}) = \frac{\sum\limits_i\,\prod\limits_{j=1}^M\,e^{-\frac{(p_j-s_i)^2
}{2\Delta_p^2}}\,\alpha_i|s_i\rangle_S}{\sqrt{\sum\limits_i\,|\alpha_i|^2\prod\limits_{j=1}^M e^{-\frac{(p_j-s_i)^2}{\Delta_p^2}}}}
\label{eq:psiweakN}
\end{equation}
\subsection{Consequences for ontology}
The \emph{intrinsic randomness} of quantum theory makes no aspect of a \emph{particular realization}
predictable. For ensemble measurements the variables are \emph{independently}
distributed and the \emph{Central Limit Theorem} guarantees that as long as the number of trials is large enough, averages over even particular
realizations converge nicely to the true mean. To see what happens in the present context, where the outcomes are clearly not 
independently distributed, let us study
$y_M$, the average of M outcomes. The expectation value of $y_M$ in the joint probability distribution $P(p_1,\ldots,p_M)$
is
\begin{equation}
{\bar y}_M = \frac{1}{M}\,\int\ldots\int\,\prod\limits_{i=1}^M\,\sum_i\,p_i\,P(\{p\}) = \sum_i\,|\alpha_i|^2\,s_i
\label{eq:weakrepeatmean}
\end{equation}
Which is certainly a remarkable result.With this, the repeated weak measurements on a single copy certainly pass one critical requirement
for ontology. 
The variance in $y_M$ can likewise be calculated and it equals $\frac{\Delta_p}
{\sqrt{2M}}$. 
This makes it appear that in principle the errors can be reduced arbitrarily, reminding one of the situation in protective measurements,
except that now no restrictions need be placed on the initial states! But such appearances turn out to be highly misleading.

As argued before the crux of the ontology issue lies in the distribution function $P(y_M)$, and not just in its mean and 
variance. As shown in \cite{ndhweak}, the distribution
function $P(y_M)$ can itself be calculated explicitly. This is in spite of the outcomes not being independently distributed.
The result is
\begin{equation}
P(y_M) = \int\ldots\int\,\prod\limits_{i=1}^Mdp_i\,P(\{p\})\delta(y_M-\frac{\sum\limits_i p_i}{M})
\label{eq:weakclt}
\end{equation}
Using eqn.(\ref{eq:probweakN}), this becomes
\begin{equation}
P(y_M) = \sqrt{\frac{M}{\pi\Delta_p^2}}\,\sum\limits_i|\alpha_i|^2\,e^{-\frac{(y_M-s_i)^2M}{\Delta_p^2}}
\rightarrow \sum\limits_i\,|\alpha_i|^2\,\delta(y_M-s_i)
\label{eq:weakclt2}
\end{equation}
where we have also displayed the limiting behaviour as $M\rightarrow\infty$. 

This, as per our discussions earlier, immediately negates not just ontology but even FAPP ontology!
In other words, the distribution of $y_M$ is not only not peaked at the true average, with errors decreasing as $M^{-1/2}$, 
it is actually a weighted sum of sharp
distributions peaked around \emph{the eigenvalues}, exactly as in the strong measurement case. 
This means that averages over outcomes of a particular realization will be eigenvalues, occurring randomly but with probability $|\alpha_i|^2$.
It then follows that averages over outcomes of a particular realization do not give any information about the initial state, precisely
as in the case of the invasive strong measurements where there can clearly be no ontology! Ensemble
measurements again become inevitable. 

The other issue to be settled in this context is whether the repeated measurements on single copies are invasive or not.
It turns out that a very large number of repeated weak measurements on a single copy has the same invasive effect as a strong measurement.
This can be seen by examining the
expectation value of the system reduced density matrix, $\rho_>^{rep}$: 
\begin{equation}
\rho_>^{rep} = \rho -\sum\limits_{i,j}\,\alpha_i\alpha_j^*\,(1-e^{-\frac{M(s_i-s_j)^2}{4\Delta_p^2}})|s_i\rangle\langle s_j|
\label{eq:weakreprho}
\end{equation}
It is seen
that as M gets larger and larger, there is significant change in the system state. In the limit $M\rightarrow \infty$, the off-diagonal
parts of the density matrix get completely quenched, as in decoherence, and the density matrix takes the diagonal form in the eigenstate of S
basis:
\begin{equation}
\rho_>^{rep}\rightarrow \sum\limits_i\,|\alpha_i|^2|s_i\rangle\langle s_i|
\label{eq:weakrholimit}
\end{equation}
which is exactly the post-measurement density matrix in the case of a strong measurement! 
The sequence of system states of eqn.(\ref{eq:psiweakN}) is a \emph{random walk} on the state space of the system(see also \cite{korotkovprb60}). It follows from 
eqn.(\ref{eq:psiweakpost}) that the eigenstates of S are the \emph{fixed points} of the probabilistic map that generates this walk.
Presumably each walk terminates in one of the eigenstates but which eigenstate it terminates in is unpredictable. 
\subsection{Other equivalent results}
\label{sec:partialonto}
Alter and Yamomoto have obtained a number of very significant results about the possibility of obtaining information about single
uantum systems \cite{alterbook,alterzeno,alterqnd}. In particular they also gave an analysis based on joint and conditional probabilities
applied to \emph{repeated weak QND} measurements on a single state \cite{alterqnd}. They too obtained evolutions resembling random walks in state
space. They concluded that it is not possible to obtain any information on unknown single states from the statistics of repeated
measurements. The degradation of the state and relation to projective measurements were not explicitly studied. In another work, they found connections between \emph{Quantum Zeno Effect} and the problem of repeated measurements
and again concluded that it is impossible to determine the quantum state of a single system. Our results on information cloning and the
general results from optimal cloning discussed in the next two sections that it may be possible to obtain partial results.

In a very interesting approach to these ontological questions, Paraoanu has investigated these issues within what he calls \emph{partial measurements} \cite{sorinpartial,sorinrepeat}. By employing
a combination of repeated such measurements on a single state and the possibility of reversing such measurements, he too has concluded
the impossibility of obtaining any information about single unknown states. The invasive aspects as well as the connections to strong measurements are not explored here either.
\section{Information Cloning and Ontology}
\label{sec:infoonto}
{\label{intro}}

As we saw in Sec.(\ref{sec:nocloneonto}), a subtle inner consistency of quantum theory prevents determining
the unknown state of a single copy by trying to make many clones of it. We had, however, proposed what we
called \emph{information cloning} in \cite{incl:2002}. The main idea was to make many copies of an unknown state
which are however not identical to the original state, but contain the same amount of \emph{information}
as the original. Now we discuss the implications for ontology of such a cloning scheme \cite{statsingle}.

The details of how this type of cloning can be used
to 
determine the state of a
{\it single unknown coherent state of quantum harmonic oscillators} 
can be found in \cite{statsingle}.
In the case of coherent states of harmonic oscillators(say, in one dimension), complete
information about the state is contained in a single complex coherency
parameter $\alpha$. Thus by information cloning what we mean is the
ability to make arbitrary number of copies of coherent states 
whose coherency parameter is $c(N)\alpha$ where $\alpha$ is the coherency
parameter of the unknown coherent state and $c(N)$ is a known constant
depending on the number of copies made.

To this end consider $1+N$ systems of harmonic oscillators whose
creation and annihilation operators are the set $(a,a^\dag),
(b_k,b_k^\dag)$ (where the index $k$ takes on values $1,..,N$).
The ${\mathrm a}$ oscillators represent the original unknown state,
and the ${\mathrm b}$ oscillators represent the information clones.
These operators satisfy the commutation relations
\beeq 
\label{eq:icloneosc}
[a,a^\dag]~=~1;~~[b_j,b_k^\dag]~=~\delta_{jk};~~[a,b_k]~=~0;~~[a^\dag,b_k]~=~0
\eneq
Coherent states parametrised by the complex number $\alpha$ are given by
\beeq 
\label{eq:iclonecoh}
|\alpha >~=~D(\alpha)~|0 >
\eneq
where $|0>$ is the ground state and the unitary operator $D(\alpha)$ is given by
\beeq 
\label{eq:icloneD}
D(\alpha)~=~e^{\alpha~a^\dag~-\alpha^*~a}
\eneq
We view the information cloning to be a unitary process. The initial composite state can be taken to be 
a {\it disentangled} 
state 
containing the unknown initial coherent state and some \emph{known} states of the ${\mathrm b}$-oscillators. 
It turns out to be best to take them also to be coherent states. In other words, the state before information cloning is
taken as
$|\alpha >|\beta_1 >_1|\beta_2 >_2...|\beta_N >_N$, where $\alpha$ is {\em unknown}
while $\beta_i$ are {\em known} to very high accuracy.
Consider the action of the unitary transformation
\beeq 
\label{eq:iclonebeam}
U = e~~^ {~t(a^\dag\otimes\sum_j r_j b_j - a\otimes\sum_j r_j b_j^\dag)}
\eneq
The most general unitary transformation of this type would 
involve complex$r_j$'s. But this can be reduced to 
the present form through suitable redefinitions of the phases of the creation and annihilation operators.
\cite{incl:2002}. Of course, such redefinitions should maintain the algebra of eqn.(\ref{eq:icloneosc}).
The process implemented by this unitary transformation is well known in Optics and is called the \emph{Beam Splitter}.
But it is very important to appreciate that we are dealing with here is when this acts on a \emph{single photon} state,
a circumstance in which the notion of a \emph{beam} is neither meaningful nor useful.
By an application of the Baker-Campbell-Hausdorff 
identity and the fact that 
$U|0>|0>_1..|0>_N=|0>|0>_1..|0>_N$ it follows
that the resulting state is also a \emph{disentangled} 
set of coherent states expressed by
\beeq 
\label{eq:iclonetrans}
|\alpha^\prime>|\beta_1^\prime>_1..|\beta_N^\prime>_N~~
=~~U~~ |\alpha>|\beta_1>_1..|\beta_N>_N~~
\eneq
In other words, the unitary transformation U acting on various coherent states induces another unitary transformation ${\cal U}$
among 
the coherency parameters. Details can be found in \cite{incl:2002}; we merely give the final result and discuss its physical
implications.
Let us Define
\beeq 
\label{eq:newalphas}
 a(t)~=~ U~ a~ U^\dag~~~~~~~ b_j(t)~=~ U~ b_j~ U^\dag
\eneq
The explicit form of the transformation induced on the parameters
$(\alpha,\beta_j)$ can be represented by the
matrix $ {\cal U}$ i.e 
\begin{equation}
\label{eq:inducedU}
{ \alpha}_a(t)= {\cal U}_{ab}{ \alpha}_b. 
\end{equation}
where we have introduced 
the notation 
$\alpha_a$ with $a=1,...,N+1$
such that 
\beeq
\label{eq:redifalpha}
\alpha_1 = \alpha  \quad\quad\alpha_k=\beta_{k-1} k\geq 2
\eneq 
Then we have
\beeq 
\label{eq:icloneind}
{\cal U}_{1a} = \left(\begin{array}{ccccc}
                             {\cos Rt}&{r_1\over R} {\sin Rt}
			     & ..&..&{r_N\over R} {\sin Rt}
			     \end{array}\right)
\eneq
where $R=\sqrt(\sum_j~r_j^2)$ and 
\beeq 
\label{eq:icloneind2}
{\cal U}_{ab} = -{r_{a-1}\over
R}~{\sin Rt}~\delta_{b1}+(1-\delta_{b1}) M_{a-1,b-1}
\eneq
where eqn (\ref{eq:icloneind2}) is defined for $a\geq 2$. 
Equivalently
 \begin{equation} \label{21}
         {\cal U} = \left( \begin{array}{ccccc}
 {\cos Rt}&{r_1\over R}~{\sin Rt}  &..  &..  &{r_N\over R}~{\sin Rt}\\
 -{r_1\over R}~{\sin Rt}& M_{11}&.. &.. & M_{1N}\\
 .. &.. &.. &.. &\\
 .. &.. &.. &.. &\\
 -{r_N\over R}~{\sin Rt}& M_{N1}&.. &.. & M_{NN}\\

	     \end{array} \right )
 \end{equation}

It is best to choose $\{\beta_i,r_i\}$ in such a way that
all $\beta_i(t)$ become identical and we get N identical
copies. This happens only when $r_i = r, \beta_i = \beta$.
In that case we have
\beeq
\beta_i(t) = -{\alpha\over{\sqrt N}}~\sin Rt + \beta 
~\cos Rt
\eneq
There is still the freedom to choose Rt. Let us first consider the choice of ${\sin Rt}=-1$ which gives N copies 
of the state $|{\alpha\over {\sqrt N}}\rangle$. This is
what we called {\em information cloning} in \cite{incl:2002}
as the states $|{\alpha\over{\sqrt N}}\rangle$ and
$|\alpha\rangle$ have the same information content. This particular choice of
$Rt$ will be seen to be optimal in the sense that it gives the least variance
in the estimation of $\alpha$. In this case the value of $\beta$ is immaterial.

It is easily seen that eqn.(\ref{eq:cloneinner}) does not pose any difficulties for information cloning, as
it did for universal cloning!
Now we can address the ontology issue by attempting to use the N-information clones for an ensemble determination
of the information-clone state first, and then a state determination of the original unknown state subsequently, by using
the fact that the information clone has the same information content as the original. More specifically, we can now use the $N$ copies of $|{\alpha\over\sqrt N}$
to make {\it ensemble measurements} to estimate ${\alpha\over\sqrt N}$ and consequently $\alpha$. 

One can already sense some limitations of the method: usually, the statistical errors can be made arbitrarily small by making
the ensemble size larger and larger.
However, in our proposal
even though the number of copies $N$ can be
made arbitrarily {\em large}, at least in principle, the coherency parameter given by 
${\alpha\over\sqrt N}$ becomes {\em arbitrarily small} while the
{\em uncertainties} in $\alpha$, being characterstic of coherent states, remain the same as in the original state.
We now address the question as to how best the original state can be reconstructed.

On introducing the {\em Hermitean} momentum and position operators $\hat p,\hat x$ through
\beeq \label{43}
{\hat x} = {(a+a^{\dag})\over\sqrt 2}~~~~ {\hat p} = {(a-a^{\dag})\over\sqrt 2i}
\eneq
the {\em probability distributions } for position and 
momentum in the coherent state $|{\alpha\over\sqrt N}\rangle$ are given by
\beeqar \label{dist}
|\psi_{clone}(x)|^2 &=& {1\over\sqrt\pi} e^{-(x-\sqrt {2\over N} \alpha_R)^2}\nonumber\\ 
|\psi_{clone}(p)|^2 &=& {1\over\sqrt\pi} e^{-(p-\sqrt {2\over N} \alpha_I)^2} 
\eneqar
Let us distribute our $N$-copies into two groups of 
$N/2$ each and use one
to estimate $\alpha_R$ through position measurements 
and the other to
estimate $\alpha_I$ through momentum measurements.
Let $y_N$ denote the average value of the position
obtained in $N/2$ measurements and let $z_N$ denote the
average value of momentum also obtained in $N/2$
measurements. The {\em central limit theorem} states
that the probability distributions for $y_N,z_N$ are
given by
\beeqar 
\label{eq:iclonecentral}
f_x(y_N)&=&\sqrt{N\over 2\pi}e^{-{N\over 2}(y_N-\sqrt{2\over N}\alpha_R)^2}\nonumber\\
f_p(z_N)&=&\sqrt{N\over 2\pi}e^{-{N\over 2}(z_N-\sqrt{2\over N}\alpha_I)^2}
\eneqar
It is more instructive to recast these as the probability distributions for ${\bar\alpha}_{R,N}, {\bar\alpha}_{I,N}$, 
the average over N measurements of
$\alpha_R, \alpha_I$:
\beeqar 
\label{eq:icloneont}
f_R({\bar\alpha}_{R,N})&=&\frac{1}{\sqrt{\pi}}\,e^{-({\bar\alpha}_{R,N}-\alpha_R)^2}\nonumber\\
f_I({\bar\alpha}_{I,N})&=&\frac{1}{\sqrt{\pi}}\,e^{-({\bar\alpha}_{I,N}-\alpha_I)^2}
\eneqar
Thus the original unknown $\alpha$ is correctly estimated, in the sense that the above distributions peak precisely
at the coherency parameter $\alpha$ of the original state. But this is not enough
and one needs to know the reliability of this estimate. For that one needs the variances.
The variances for $\alpha_N$ are easily found out from eqn.(\ref{eq:icloneont}):
\beeq
\label{eq:iclonevar}
\Delta \alpha_{R,N} = \Delta \alpha_{I,N} = \frac{1}{\sqrt{2}}
\eneq
Thus, while the statistical error in usual measurements goes as $\frac{1}{\sqrt{N}}$,
and can be made arbitrarily small by making $N$ large enough, information cloning gives
an error that is fixed and equal to the variance associated with the original unknown state.
For coherent states with large enough $\alpha$, even these errors are quite reasonable. Another figure of merit,
the so called \emph{Fidelity} has also been adopted in \cite{incl:2002,optimal1,optimal2,optimal3,fidel1,fidel2}.
That fidelity for information cloning works out to 1/2 \cite{incl:2002}, the maximum possible for the so called
\emph{Gaussian Cloning} \cite{optimal1,optimal2,optimal3,fidel1,fidel2,rmp77,lindblad2000,werner1999}. Therefore, 
fidelity on its own may give an
unnecessarily pessimistic picture. Comparison between information cloning and optimal clonings mentioned above
will again be made in Sec.(\ref{sec:optimcloneonto}). 


Thus we have shown that even when the coherent state is {\em unknown} single state, information cloning
will allow its determination, but with fixed statistical errors. Nevertheless, it is a great improvement
from not being able to know anything at all about the unknown state.

A comparison with our technical criteria for ontology reveals that again there is no perfect ontology but indeed there
is FAPP ontology of the first kind. In contrast, protective measurements gave a FAPP ontology of the second kind. In the 
protective case the FAPP ontology could approach perfect ontology arbitrarily close, but never equal it. In both cases,
one had to restrict the classes of states for which they would work and the restricted class did not allow linear
superpositions. 

\section{Approximate Cloning and Ontology}
\label{sec:optimcloneonto}
Though the No-cloning theorem forbids making perfect clones of an unknown state, there seems
nothing against making imperfect copies. The information cloning of the previous section was
a particularly interesting variant of this theme. Then the obvious question is the closeness
to perfect cloning that can be achieved. There has been an explosion of interest in this question
and the reader is referred to \cite{rmp77} for a comprehensive review. We shall only examine
the so called {\em optimal cloning} 
\cite{optimal1,optimal2,optimal3,fidel1,fidel2,lindblad2000,werner1999,optical,grosshans}(see \cite{rmp77} for a review), from the ontological point of view. The details are not that
critical to understanding the broad implications and chief conclusions.

In these
implementations, one starts with the {\em original 
unknown
state} $|\alpha\rangle$ belonging to the Hilbert space
${\cal H}_A$, a number of ancillary states, also known as {\em blank states},
$|b_0\rangle,|b_1\rangle...|b_N\rangle$. The ancillaries are \emph{known} states. This is the general 
setup for all cloning processes. The ancillaries belong 
to the Hilbert spaces ${\cal H}_{B_i}$; each of them is isomorphic to ${\cal H}_A$. Unlike the information 
cloning case, a number
of additional states called machine states, also known, $|m_0\rangle,|m_1\rangle.....
|m_M\rangle$ all belonging to the Hilbert spaces isomorphic to, say, ${\cal H}_M$, are also considered. The combined Hilbert space has the
structure ${\cal H}_A\otimes {\cal H}_M \otimes \prod_i
{\cal H}_{B_i}$. 

A {\it general cloning
transformation} ${\cal T}$ has the effect
\beeq
|\alpha\rangle\prod_0^N |b_i\rangle\prod_0^M |m_j\rangle
{\stackrel{\cal T}{\rightarrow}}
\sum_{i,j,k} d_{ijk}|a_i\rangle \prod_j^M |\beta_j\rangle \prod_k^N |\gamma_k\rangle
\eneq
Such a general cloning is said to be \emph{optimal} if it satisfies the two conditions: i){\it all} the reduced density 
matrices ${\bf \rho}_{i_0}$ obtained by tracing over the ${\cal H}_A$
states, the machine states and all the blank states
except those belonging to ${\cal H}_{B_{i_0}}$, are all
{\it identical} and ii) each of them has \emph{maximum} overlap with the
original unknown state $|\alpha\rangle$ i.e with the
maximum possible value of $\langle\alpha|{\bf \rho_{i_0}}|\alpha\rangle$. The reduced density matrices are in general {\em mixed}.

In the case of information cloning, the clones were all disentangled and one could use all of them at a time for carrying out
measurements of one's choice. In the case of optimal cloning, in general the clones could be in entangled states. Depending on such
details, it could even be that that at any given time it is 
possible to realise only a few of the reduced matrices ${\bf \rho}_i$ as different values of $i$ require
tracing over different states. 

As can be gathered from \cite{rmp77} and the many references there, there are various types of optimal clonings. But for the ontological
questions, only a part of them are of interest. Firstly, we need only look at the so called \emph{universal} types as these can produce
clones of unknown states. The so called \emph{state dependent} cloning is not of interest. The information cloning that we discussed
earlier is state dependent in one way as it can work only with coherent states, but it is also somewhat universal in the sense that the input 
state can be \emph{any} coherent state. In fact, it is a particular case of Gaussian cloning 
\cite{optimal1,optimal2,optimal3,fidel1,fidel2,rmp77,lindblad2000,werner1999}. Secondly, even among the universal optimal cloning there
are results for the so called $N\rightarrow M$ type clonings. Here N is the number of copies of the unknown initial state(usually pure)
and M the number of clones(usually mixed). For our ontological considerations, only $1\rightarrow M$ types are relevant.

Let us first consider the case where the input Hilbert space is \emph{finite dimensional}, and specifically consider only cubits. We
shall only look at a few illustrative aspects. For qubits, the fidelity F, which is the overlap of the clone with the original, is,
given by
\begin{equation}
\label{eq:optimalqubitfidelity}
F(N,M) = \frac{MN+M+N}{M(N+2)}
\end{equation}
for the $N\rightarrow\,M$ case. The clone state is of the form (with $tr\,\rho\cdot\rho^{\perp}=0$),
\begin{equation}
\label{eq:clonequbit}
\rho^{clone} = F\,\rho^{ini}+(1-F)\,\rho^{\perp}
\end{equation}
The accuracy of the state determination with the clones requires as large a M as possible. Therefore, for N=1 and $M\rightarrow\,\infty$,
one has F = 2/3. In fact, for an arbitrary M,
\begin{equation}
\label{eq:fidq1toM}
F(1,M) = \frac{2M+1}{3M}
\end{equation}
The largest value, for the non-trivial case $1\rightarrow 2$ is 5/6. But with only two clones the errors in the state determination
are very high. But as M is increased, to get more accurate state determination, F decreases, reaching the limiting value of 2/3. In
that case though the errors are very small, the estimates for expectation values of observables deviates significantly from the
true values. For example, for observables ${\cal O}$ with zero expectation values in $\rho^{\perp}$, one finds
\begin{equation}
\label{eq:qubitclonemean}
\langle {\cal O} \rangle_{clone} = \frac{2}{3}\,\langle {\cal O}\rangle_{true}
\end{equation}
failing even the first criterion for ontology rather poorly. Unlike the information cloning case, where the error was \emph{independent}
of the input state, here the error is a finite fraction(1/3) of the expectation value. The resources required even to reach this are
impractically large ($M\rightarrow\infty$).

Of course, it is inappropriate to use the results obtained for optimal cloning of qubits to make a comparison with information
cloning which is really a case of infinite-dimensional Hilbert space. But results are also available for optimal cloning for
arbitrary, but finite dimensional Hilbert space. As an intermediary to considering continuous variable cloning, let us
consider Werner's results \cite{werner1999} for d-dimensional Hilbert spaces. The formula for the fidelity of $N\rightarrow M$
cloning is
\begin{equation}
\label{eq:fiddntom}
F(N,M) = \frac{N}{M}+\frac{(M-N)(N+1)}{M(N+d)}
\end{equation}
The clone state is given by:
\begin{equation}
\label{eq:clonedntom}
\rho^{clone} = \eta(N,M)\,\rho^{ini}+(1-\eta(N,M))\,\frac{I}{d}\quad\quad \eta(N,M) = \frac{N}{M}\,\frac{M+d}{N+d}
\end{equation}
\begin{equation}
\label{eq:pqclonentom}
\rho^{clone} = \frac{N}{M}\,\rho^{ini}+(1-\eta(N,M))\,\frac{I}{d}\quad\quad \eta(N,M) = \frac{N}{M}\,\frac{M+d}{N+d}
\end{equation}
Let us look at the continuous case by letting $d\rightarrow\,\infty$ first. While the fidelity approaches the limit
$\frac{N}{M}$, the density matrix formula is much more tricky. Now if apply this formula for fidelity
to N=1, $M\rightarrow\infty$ limit relevant for our ontological concerns, we see that the fidelity vanishes!

This is because of the attempt to find a universal cloner for continuous variable case. Let us lower the expectations and consider
only coherent states. It has been shown that the fidelity is \emph{bounded} by
\begin{equation}
\label{eq:cohfidbound}
F(N,M)\le\,\frac{MN}{MN+M-N}
\end{equation}
The clone state is a mixture of coherent states centred around the unknown initial coherent state.
Its explicit form is given by (see eqn.(53) of \cite{rmp77}, but watch for a typo!):
\begin{equation}
\label{eq:optimclonestate}
\rho^{clone}(\alpha) = \frac{1}{\pi\sigma(N,M)^2}\,\int\,d^2\beta\,e^{-\frac{|\beta|^2}{\sigma(N,M)^2}}\,|\alpha+\beta\rangle
\langle \alpha+\beta|
\end{equation}
where $\sigma(N,M)$ stands for
\begin{equation}
\label{eq:optimclonesig}
\sigma(N,M)^2 = \frac{1}{N}-\frac{1}{M} \ge 0
\end{equation}
Returning to the ontology issue, we set N=1. 
It is easy to verify that the mean values of x and p in the clone state of eqn.(\ref{eq:optimclonestate}) are exactly the
 same as in the unknown original coherent state. This was so in the case of information cloning too. But the variances in
x and p for the clone state turn out to be
\begin{equation}
\label{eq:optimclonevar}
(\Delta x)^2_{clone} = \frac{1}{2}+\sigma(1,M)^2=(\Delta_p)^2_{clone}
\end{equation}
Like the information cloning case, these variances are the same for all coherent states. But irrespective of M, the variances are
worse here than there. Again there is only FAPP ontology, of a somewhat worse quality.
\subsection{Probabilistic Cloning} What was described till now can be called \emph{deterministic} cloning. There are also probabilistic
cloning machines. The reader is referred to \cite{duan1998} to get an understanding of these. Many features and implementations are
different and these cloning devices are very interesting. But from our ontological perspective, the situation is not too different; again
the mean values can approach the true values and the errors can not be completely eliminated. One can ascribe a FAPP ontology with
figures of merit determined by both of these. 
\section{Conclusions}
\label{sec:conclusions}
In this paper we have carefully examined the issue of obtaining information about the state of a single quantum system. We have
equated the ability to obtain such information with the concept of \emph{ontology} in quantum mechanics. We have given a precise
technical characterisation of this concept and examined the implications of a large variety of quantum measurements including
projective measurements, protective measurements, weak measurements(including weak QND measurements) and the so called partial
measurements. We have also examined the issue in the light of the no-cloning theorem on the one hand, and in the light of a variety 
of cloning techniques. 

The impossibility of gaining information about a single quantum state is considered to be the basic tenet of quantum mechanics. Admittedly,
it was based on the picture of quantum measurements that dominated during the early development of quantum theory. Central to that line
of thinking were the highly invasive nature of the eigenvalue-eigenstate based projective measurements. In view of the highly invasive
nature of such measurements, that thinking seemed almost obvious. But what is surprising now is that when even novel measurement
schemes like weak measurements, partial measurements are around, which make such a tenet far from obvious, it still remains rock solid.
Now the results that even these seemingly non-invasive measurement schemes simply can not coax any information out of generic single states
make this lack of ontology deep and perplexing, as if they are the foundational principles of quantum theory. Nevertheless, that schemes
like protective measurements, information cloning in particular and optimal cloning in general exist to provide a silver lining
in the form of what we have called FAPP ontology
is also equally perplexing. What general principles are lurking behind these is something that all those trying to fathom the depths of quantum
theory will be eagerly searching for.


\end{document}